\documentstyle[preprint,aps,epsfig]{revtex}

\newcommand{\beq}{\begin{eqnarray}}
\newcommand{\eeq}{\end{eqnarray}}
\newcommand{\beqs}{\begin{eqnarray}}
\newcommand{\eeqs}{\end{eqnarray}}
\newcommand{\bary}{\begin{array}}
\newcommand{\eary}{\end{array}}

\newcommand{\figpos}{p}      
\newcommand{\spaAR}{6.5}     
\newcommand{\heightAR}{14.0} 
\newcommand{\widthAR}{14.0}  

\newcommand{\cL}{{\cal L}}

\def\eq#1{{\rm eq.}~(\ref{#1})}

\def\eps{\varepsilon}

\def\Oc{{\cal O}}

\def\eV{{\rm~eV}}

\def\GeV{{\rm~GeV}}
\def\TeV{{\rm~TeV}}

\def\et{{\it et. al}}

\def\npa#1{Nucl.\ Phys.\ {\bf A #1}}

\def\plb#1{Phys.\ Lett.\ {\bf B #1}}

\def\prd#1{Phys.\ Rev.\ {\bf D #1}}
\def\prl#1{Phys.\ Rev.\ Lett. {\bf #1}}

\def\epjc#1{Eur.~Phys.~J.\ {\bf C #1}}

\def\pha#1{{\tt hep-ph/#1}}
\def\phb#1{{[{\tt hep-ph/#1}]}}
\def\tha#1{{\tt hep-th/#1}}
\def\thb#1{{[{\tt hep-th/#1}]}}

\def\exb#1{{[{\tt hep-ex/#1}]}}
\def\aspha#1{{\tt astro-ph/#1}}
\def\asphb#1{{[{\tt astro-ph/#1}]}}

\def\nucexb#1{{[{\tt nucl-ex/#1}]}}

\def\gsim{\ \rlap{\raise 3pt \hbox{$>$}}{\lower 3pt \hbox{$\sim$}}\ }
\def\lsim{\ \rlap{\raise 3pt \hbox{$<$}}{\lower 3pt \hbox{$\sim$}}\ }

\def\overset#1#2{\ \rlap{\raise 3pt \hbox{$#1$}}{\lower 1pt\hbox{$#2$}}\ }
\def\underset#1#2{\ \rlap{\lower 3pt \hbox{$#1$}}{\raise 2pt \hbox{$#2$}}\ }

\def\spur#1{\mathord{\not\mathrel{#1}}}

\def\putFig#1#2#3#4#5#6#7 
{
 \newpage
 \begin{figure}[\figpos]
 \LARGE
 $#7$
 \begin{center}
 \mbox{\epsfig{figure=#1.eps,angle=0,width=#3cm,height=#4cm}}
 $#6$
 \end{center}
 \vspace{1.5cm}
 \caption{#2}
 \label{#1}
 \end{figure}
}

\begin{document}

\preprint{\vbox{\hbox{WIS/34/02-AUG-DPP}
                \hbox{hep-ph/0208102}}}

\title{Lepton Masses and Mixing in a Left-Right Symmetric Model with a
TeV-scale Gravity}

\author{Gilad Perez  \\
        \small \it Department of Particle Physics,
        Weizmann Institute of Science,
        Rehovot 76100, Israel}
\maketitle

\begin{abstract}
We construct a left-right symmetric (LRS) model in five dimensions
which accounts naturally for the lepton flavor parameters.
The fifth dimension is described by an orbifold, $S_1/Z_2\times Z'_2$,
with a typical size of order TeV$^{-1}$.
The fundamental scale is of order 25 TeV
which implies that the gauge hierarchy problem is ameliorated.
In addition the LRS breaking scale is of order few TeV
which implies that interactions beyond those of the standard model
are accessible to near future experiments.
Leptons of different representations are localized around different
orbifold fixed points. This explains, through the Arkani-Hamed-Schmaltz
mechanism,
the smallness of the tau mass compared to the electroweak breaking scale.
An additional U(1) horizontal symmetry, broken by
small parameters, yields the hierarchy in the
charged lepton masses, strong suppression of the light neutrino masses
and accounts for the mixing parameters.
The model yields several unique predictions.
In particular, the branching ratio for the lepton
flavor violating process $\mu^-\to e^+ e^- e^- $ is comparable with its
present experimental sensitivity.

 \end{abstract}

\section{Introduction}
The recent results from the SNO~\cite{SNO} and other~\cite{Nuexp} experiments
provide strong evidences 
for the incompleteness of the standard model (SM).
Among the various new physics (NP) scenarios that predict neutrino masses, the
left-right symmetric (LRS) framework~\cite{LRSM} is 
an attractive and popular one.

In many LRS models, {\it e.g.} models embedded in GUT~\cite{GUT},
the LRS breaking scale, $v_R\gsim10^{14}\GeV$, 
is much higher than the electroweak (EW)
breaking scale, $k_1\sim10^{2}\GeV$, and the low energy effective
Lagrangian is similar in many aspects to that of the SM. 
In such a case present and near future experiments will not be able 
to directly probe the NP. 
Furthermore, the introduction of such a high scale, in addition to
the Planck scale, $M_{P}\sim 10^{19}\GeV$, 
related to gravity in four dimensions, raises the
gauge hierarchy and fine tuning problems shared by many models.
  
A new exciting possibility, however, that the
fundamental scale of gravity can in fact be much smaller than $M_{P}$
was raised in~\cite{LED}. 
It is very interesting, therefore, to 
investigate whether a natural LRS model (LRSM), in which 
both $M_P$ and $v_R$ are low, say below 100 TeV,
and in which the neutrinos are very light, can be constructed.

This work is focused on the lepton flavor parameters, we comment on
the inclusion of the quark sector in the conclusion.
We present a LRSM which naturally accounts for the
flavor parameters of the lepton sector and in which the fundamental scale
and the LRS breaking scale are of the order of or below 25 TeV.
It is a model of, at least,~\footnote{To have $M_P\lsim
100\TeV$ requires more extra dimensions.} 
one extra compact dimension
which copes with the above problems by using both
the Froggatt-Nielsen~\cite{Fro} and the Arkani-Hamed-Schmaltz (AS)~\cite{HS}
mechanisms.
The main role of the AS mechanism is to
localize ``left'' and ``right'' lepton fields on  different
fixed points in the extra dimension, which then explains the smallness of
$m_\tau$ compared with the electroweak breaking scale. An
additional U(1) horizontal symmetry, described in detail
in~\cite{KP}, yields a modified see-saw mechanism and accounts
for the other flavor parameters.

Before proceeding with the details of our work we note that
recent two papers, by Mimura and Nandi~\cite{MN} and by
Mohapatra and Perez-Lorenzana~\cite{MP}, dealt with
the construction of a LRSM in five dimensions [5D].
Though important parts of the analysis in~\cite{MN,MP} are general and
also used below the three
models  use, in fact, different 
constructions in the extra dimension.
Furthermore, to the best of our knowledge,
the present model is the only LRSM in 5D that aims to account 
naturally for all the lepton flavor parameters.

To make our discussion concrete we 
briefly list the lepton flavor parameters
as deduced from various experiments.
The charged lepton masses are~\cite{PDG},
\beq
m_e\simeq5.1\cdot10^{-1}~{\rm MeV}\,, \ \
m_\mu\simeq1.1\cdot10^{2}~{\rm MeV}\,,\ \
m_\tau\simeq1.8\cdot10^{3}~{\rm MeV}\,.
\label{mcl}
\eeq
As for the neutrino parameters, we consider
below the large mixing angle solution of the solar
neutrino problem  which is favored by data~\cite{Analysis,Rev1}. 
Consequently, the
neutrino mass differences, at the 3$\sigma$ CL, are:
\beq
\Delta m^2_{\rm Sol}&=&(0.2-3)\cdot10^{-4}\eV^2 \ ,\ \ 
\Delta m^2_{\rm Atm}=(1-6)\cdot10^{-3}\eV^2\,,
\label{numass}
\eeq
where $\Delta m^2_{\rm Sol}$ [$\Delta m^2_{\rm Atm}$] 
is the mass square difference deduced from the data of
the solar [atmospheric] neutrino experiments. 
The neutrino mixing parameters are~\cite{Analysis,Rev1}:
\beq
\tan^2\theta_{12}=0.2-0.7\ ,\ \
\tan^2\theta_{23}=0.4-3\ ,\ \
\theta_{13}\lsim0.15\,.
\label{numix}
\eeq
In addition there are both direct~\cite{PDG,Beta,Pet1} and 
indirect~\cite{Cos} bounds on
the absolute scale of the neutrino masses:
\beq
m_i&\lsim& 1 \eV\,.
\label{dbd1}
\eeq

In section \ref{Model}
we present our 5D LRSM model. 
In section \ref{4D} we describe the 4D effective
theory of our model and calculate its lepton flavor parameters.
In section \ref{Test} we suggest several criteria by which we can  
test the model predictions. 
Comments and conclusions are given in section \ref{Conclusion}.


\section{The model}\label{Model}


The space time of our model is described by the usual 4D space and 
an additional space dimension compactified on the orbifold
$S_1/Z_2\times Z'_2$. 
The characteristic energy scales are:
 \beq \ L^{-1}\sim \TeV \,; \qquad
 L^{-1}<v_R, v\lsim M_*\,; \qquad
a\equiv v\sqrt{ L  \over  M_*}\gsim 5 \label{scales}\,,
 \eeq
where $L$ is the size of the fifth dimension fundamental domain,
$v_R$ is the scale at which the LRS is broken spontaneously,
$v$  is related to the typical width of the fermion wave functions and
$M_*$ is the fundamental scale.

The discrete group, $Z_2\times Z'_2$ yields the following
identifications for the fifth dimension coordinate, $y$:
\beq\label{orbif}
Z_2:\, \ y\leftrightarrow -y\,; \qquad Z'_2:\, \ y'\leftrightarrow -y'
\,,
\eeq
where $y'\equiv y + \pi R/2$.

The symmetry of the model is given by,~\footnote{To avoid confusion when dealing
with fermions in 5D we denote the two SU(2) gauge groups as
SU(2)$_{1,2}$. We switch back to the ordinary notations, SU(2)$_{L,R}$, 
when we discuss the 4D effective theory.}
\beq
{\rm SU(2)}_1\times{\rm SU(2)}_2
\times{\rm U(1)}_{\rm B-L}\times{\rm G}_{\rm LR}\times{\rm U(1)}_{\rm H}
\,,
\eeq 
where G$_{\rm LR}$
corresponds to parity symmetry in the usual 4D LRSM
and U(1)$_{\rm H}$ corresponds the global horizontal symmetry~\cite{KP}. 
The gauge group is broken both explicitly,
by the transformation laws of the fields under the orbifold discrete
group, and spontaneously, by the VEVs of the scalars~\cite{MN}.

We now move to describe the field content of our model.
We first describe the scalar sector, then we  move to the lepton
sector and afterwards to the gauge boson sector.
For each sector we describe the transformation of the fields 
under the gauge and the orbifold groups while the horizontal 
charges are given in the next section.  
 
The scalar field content of the model is similar to the minimal
LRSM (see e.g. \cite{LRSM,KP,Gun1}). $\phi_{1}$ is
a bi-fundamental of the two SU(2) groups and $\Delta_{1,2}$ are
triplets 
of the SU(2)$_{1,2}$ gauge groups:
\beq
\phi_1=\pmatrix{h^0_1&h_2^+\cr h_1^-&h^0_2}\,,\qquad 
\phi_2\equiv\tau_2\,\phi_1^*\, \tau_2\,\,,\qquad 
\Delta_{1,2}=\pmatrix{{\delta^+_{1,2}\over\sqrt2}&\delta^{++}_{1,2}\cr 
\delta^0_{1,2}&-{\delta^+_{1,2}\over\sqrt2}}
\,.
\eeq 
In addition, we assume the existence of
a real scalar field $\varphi$, a pseudo-singlet of the discrete
LRS group, 
\beq
\varphi\,{\overset{^{\,{\rm G}_{\rm LR}}}{\longleftrightarrow}}   \,-\varphi
\label{GLR}\,,
\eeq
where its self interactions and coupling to the   
fermions are discussed in the appendix.

The transformation laws of the scalars under the orbifold discrete group
are given by,
\beq
\phi_1(x^\mu, -y) &=& \phi_1(x^\mu, y) \,, \qquad
\phi_1(x^\mu, -y^\prime) = \phi_1(x^\mu, y^\prime) P' \,;
\nonumber  \\
\Delta_2(x^\mu, -y) &=& \Delta_2(x^\mu, y) \,, \qquad
\Delta_2(x^\mu, -y^\prime) = -P'\Delta_2(x^\mu, y^\prime) P' \,;
\nonumber  \\
\Delta_1(x^\mu, -y) &=& \Delta_1(x^\mu, y) \,, \qquad
\Delta_1(x^\mu, -y^\prime) =\Delta_1(x^\mu, y^\prime)\nonumber  \\
\varphi(x^\mu, -y) &=&-\varphi(x^\mu, y)\,, \qquad 
\varphi(x^\mu, -y') =-\varphi(x^\mu, y')
\,, \label{transcal}
\eeq
with $P^\prime = {\rm diag}(1,-1)$.
Note that, as already discussed in~\cite{MN}, only one of the
neutral components of $\phi_1$, $h^0_1$, has a zero mode and
can develop a VEV. This fact is related to the natural suppression of the
Dirac neutrino masses, which is necessary for the
phenomenological viability of our model~\cite{KP}.

The field content of the lepton sector is more involved.
It is similar to the one of~\cite{MP} but not identical since our
mechanism of generating neutrino masses is very different from the one
of~\cite{MP}. 
In most of the LRS models there is a pair of lepton doublets (connected by
G$_{\rm LR}$) for each generation.
In our model, we actually
introduce two such pairs for each generations (see also~\cite{MP}).
The reason for this is that our model assumes canonical see-saw
mechanism~\cite{SeeSaw}. This implies zero modes for both the 4D
left and right handed neutrinos. The transformation law of the
bidoublet $\phi_{1}$ (\ref{transcal}) requires that the doublets
of the SU(2)$_2$ group will transform non-trivially under the
orbifold  $Z'_2$ discrete group. Thus, only one of the two
components of the doublet can have a zero mode. This enforces to
double the number of fermion fields as previously done in~\cite{MP}.
Consequently, for each generation $i$, we have four lepton doublets.
Two doublets of SU(2)$_1$ and two of SU(2)$_2$, denoted as $L^i_1$,
$L'^i_1$ and $L^i_2$,$L'^i_2$ respectively. 

The representation of the leptons under the SU(2)$_1\times$SU(2)$_2
\times$U(1)$_{\rm B-L}$ gauge group is therefore given by,
\beq
L^i_1(2,1)_{-1}\, \ {\overset{^{\,{\rm G}_{\rm LR}}}{\longleftrightarrow}}  
\, \  L^i_2(1,2)_{-1}\,, \qquad
L'^i_1(2,1)_{-1}\, \ {\overset{^{\,{\rm G}_{\rm LR}}}{\longleftrightarrow}}  \, \  L'^i_2(1,2)_{-1}
\,,\label{double}
\eeq
where $i=1..3$ stands for lepton flavors.

The lepton fields have the following transformation laws
under the discrete group:
\beq 
L_1^i(x^\mu, -y) &=& - L_1^i(x^\mu, y) \,, \qquad 
L_1^i(x^\mu, -y^\prime) = - L_1^i(x^\mu,y^\prime) \,;
\nonumber  \\
L_1^{\prime i}(x^\mu, -y) &=& - L_1^{\prime i}(x^\mu, y) \,, \qquad
L_1^{\prime i}(x^\mu, -y^\prime) =L_1^{\prime i}(x^\mu, y^\prime) \,;
\nonumber  \\
L_2^i(x^\mu, -y) &=&  L_2^i(x^\mu, y) \,, \qquad
L_2^i(x^\mu, -y^\prime) = P' L_2^i(x^\mu, y^\prime) \,;
\nonumber  \\
L_2^{\prime i}(x^\mu, -y)&=&  L_2^{\prime i}(x^\mu, y) \,, \qquad 
L_2^{\prime i}(x^\mu, -y^\prime) = -P' L_2^{\prime i}(x^\mu, y^\prime) \,, 
\label{transL} \eeq
where note that unlike~\cite{MP} the transformation laws of $L_2^i$
allow for the ``right handed'' neutrinos to have zero modes.
As we shall see below, however, they acquire
large masses due to their Yukawa couplings to $\Delta_2$.
 
In order to have 
a chiral 4D low energy effective theory any lepton,
$\psi$, has, on top of (\ref{transL}), the following transformation law
under  the orbifold discrete group~\cite{GGH,Or}: 
\beq\label{chiral} \psi
(x^\mu, x_5) &\rightarrow& \psi(x^\mu, -x_5)=\gamma^5 \psi (x^\mu, x_5)
\,,\eeq
where $x_5$ stands both for $y$ and $y'$.

We now move to the gauge boson sector. 
As was already discussed in~\cite{MN,MP},
with the above transformation laws for the matter fields,
the gauge bosons of the SU(2)$_2\times$U(1)$_{B-L}$ gauge groups
have the following transformations:
\begin{eqnarray}
&&W_{\mu}(x^{\mu}, -y)=W_{\mu}(x^{\mu},y) \,, \qquad
W_{\mu}(x^{\mu},-y')=P'W_{\mu}(x^{\mu},y') P'\,; \nonumber \\
&&W_{5}(x^{\mu}, -y)=-W_{5}(x^{\mu},y) \,, \qquad
W_{5}(x^{\mu}, -y')=-P'W_{5}(x^{\mu},y)  P'\,;\nonumber \\
&&B_\mu(x^\mu,-y),B_\mu(x^\mu,-y')
=B_\mu(x^\mu,y),B_\mu(x^\mu,y')\,;\nonumber \\
&& B_5(x^\mu,-y),B_5(x^\mu,-y')
=-B_5(x^\mu,y),-B_5(x^\mu,y')\,,
\label{W2B}
\end{eqnarray}
where $B_{\mu,5}$ corresponds to the U(1)$_{B-L}$ gauge group.
The gauge bosons of the SU(2)$_1$
have the following transformations:
\begin{eqnarray}
W_{\mu}(x^{\mu}, -x_5)=W_{\mu}(x^{\mu},x_5) \,, \qquad
W_{5}(x^{\mu}, -x_5)=-W_{5}(x^{\mu},x_5) \,,
\label{W1}
\end{eqnarray}
where $x_5$ stands both for $y$ and $y'$.

The additional U(1)$_{\rm H}$ symmetry (discussed in detail in \cite{KP}),
is assumed to be broken by small parameters in two stages. First
it is broken to a discrete $Z_2$ symmetry (the discrete symmetry
does not allow for Majorana masses) by a small parameter, $\eps$.
Then the discrete subgroup is further broken by a small parameter,
$\delta$. Thus, as discussed below, various terms in
the 5D effective Lagrangian are suppressed by powers of $\eps$ and
$\delta$: 
\beq {\cal L}^{5D}&=& {\cal L}_0+{1\over\sqrt {M_* }}
\left[  {\eps}^{| Q(L^j_2)-Q(L^i_1)+Q(\phi_1) |} f \bar L^i_1
\phi_1 L^j_2 +  {\eps}^{| Q(L^{j\prime}_2)-Q(L^i_1)-Q(\phi_1) |}g
\bar L^i_1 \phi_2 L^{j\prime}_2\right.
\nonumber \\
&+& \left. \delta{\eps}^{| Q(\Delta_1)+Q(L^i_1)
+Q(L^j_1)\pm{1\over2}|}h({L_1^i}^T i\tau_2 \Delta_1 L^j_1
+{L^i_2}^T i\tau_2 \Delta_2 L^j_2)+L^i_a\leftrightarrow
L_a^{i\prime}\right] \nonumber \\
&+&V^{5D}(\phi,\Delta) \label{Lcorr} \,, \eeq
where ${\cal L}_0$ contains the kinetic and the gauge interaction
terms~\cite{MN,MP}.
Higher dimensional operators are subdominant
for the low energy effective theory. This is due to suppression factors
coming from inverse powers of $M_*\sim 25\TeV$ and from 
powers of $\epsilon$ and $\delta$~\cite{Ar}.~\footnote{As an example consider the
operator $L_1^* L_1^*  \phi_2\Delta_2\phi^\dagger_1$ which is similar to the
operator that gives the dominant contribution 
for neutrino masses in~\cite{MP}. In our model the induced neutrino 
mass from such an operator is roughly,  
$\eps^{24}\delta k_1^2 v_R / M^2_*$, which is completely negligible.} 
Furthermore, we 
assume that quantum modifications to our model both from perturbative 
and non-perturbative sources are small in the IR limit of the 4D effective
theory (for discussions on this subjects see e.g~\cite{Loops,Anom} 
and references therein).

The effective 5D scalar potential, $V^{5D}(\phi,\Delta)$ is given by
\beq V^{5D}(\phi,\Delta) & = & \sum_{ijk}\left\{ -\mu^2_{ii}
\mbox{Tr}(\phi^\dagger_i \phi_i ) -\mu^2_i \mbox{Tr}(\Delta_i
\Delta_i^\dagger )+ {1\over M_* } \left[\lambda_{ij}   \mbox{Tr}(
\phi^\dagger_i \phi_j )
 \mbox{Tr}(\phi^\dagger_i  \phi_j)\right.\right.\nonumber\\
& + & \rho_{ij}  \mbox{Tr} (\Delta_i \Delta_i )
\mbox{Tr} (\Delta_j^\dagger \Delta_j^\dagger )
+\left.\left.
\alpha_{ii}  \mbox{Tr} (\phi^\dagger_i \phi_i )
\mbox{Tr} (\Delta_k \Delta_k^\dagger )
+ \beta_{ii}\mbox{Tr}(\Delta_1^\dagger \Delta_1\phi_i \phi^\dagger_i
\right.\right.\nonumber\\
&+&\left.\left.
\Delta_2^\dagger \Delta_2\phi^\dagger_i \phi_i ) 
+ \gamma_{21} \mbox{Tr}(\Delta_1^\dagger
\phi_2\Delta_2\phi^\dagger_1) \right]\right\} \,, \label{VDelphi}
\eeq where some of the coefficients in $V^{5D}$ contain
implicit suppression factor of various powers of $\eps$~\cite{KP}.
In the generic case, $\phi_{1.2}$ and $\Delta_{1,2}$  
develop VEVs,
\beq
\langle \phi_{1} \rangle =\sqrt M_* \pmatrix{
                             k_1 & 0 \cr
                             0& 0}\,, \qquad
 \langle \Delta_{1,2} \rangle &=&\sqrt M_* \pmatrix{
                             0 & 0 \cr
                             v_{L,R}e^{i\alpha_{L,R}} & 0}\,. 
\label{vevs}
\eeq
The different VEVs have the following relation among them~\cite{LRSM,Gun1},
\beq
|v_R v_L|=\gamma |k_1|^2
\,.
\label{rel}
\eeq
In the presence of the U(1)$_{\rm H}$ $\gamma$ is given by~\cite{KP},
\beq
\gamma \sim  \eps^{|Q(\Delta_R)-Q(\Delta_L)-2Q(\phi_1)|}
\label{gammasup}
\,.
\eeq

\section{The Spectrum of the 4D Theory}\label{4D}
Most of the details of the model were given above. 
To calculate
its low energy spectrum, however, two main ingredients are missing.
One is the horizontal charges of the fields, which are given below. The other
is related to the  separation and localization of the different
lepton fields. 
We assume that all the doublets of the SU(2)$_1$ gauge group have
same-sign Yukawa couplings to $\varphi$ (\ref{L5}).
Since we choose $\varphi$ to be odd under the LRS discrete
group (\ref{GLR}), all the SU(2)$_2$ doublets have the opposite 
sign of Yukawa couplings to 
$\varphi$.~\footnote{In that way we go one step further in eliminating the
arbitrariness in the sign assignment of the corresponding Yukawas. Such
a problem is often encountered in the AS framework.}

In the model presented below, we assume that 
these Yukawa couplings are flavor independent.  
Consequently, the magnitude of the couplings between the leptons and $\varphi$ 
is universal and naturally of order unity.
This assumption can be motivated in cases where the couplings to the
scalar are originated from a flavor blind sector of
a more fundamental theory. We shall comment on the implications
of relaxing this assumption in the following section.

As explained, the fields $L^i_{1}$ and $L^j_{2}$ 
are localized around different fixed points~\cite{GGH,Kap}.
In the appendix we show that
bidoublet Yukawa couplings are
naturally suppressed due to small overlap between the wave functions of
the zero modes of $L^i_{1}$ and $L^j_{2}$.
As shown in (\ref{K}), given the model fundamental parameters (\ref{scales}),
the corresponding suppression
factor, $K\sim e^{-a}$, is naturally of the order of $m_\tau/k_1$.

At this stage, when our main focus is on the zero mode of the fields, it is
convenient to switch to the ordinary 4D notations, 
$SU(2)_{1,2}\rightarrow SU(2)_{L,R}\,; \ $ 
$\Delta_{1,2}\rightarrow \Delta_{L,R}\,; \ $
$f^0_{1,2}\rightarrow f^0_{L,R}\,,$
where $f^0_i$ is any fermion zero mode that belongs to a doublet of the SU(2)$_i$
gauge group.

The charges of the fields under the U(1)$_{\rm H}$ horizontal symmetry
are given by:
\beq
Q(L_L^3)&=&0\,, \qquad Q(L_L^2)=1\,,\qquad Q(L_L^1)=3\,, \nonumber\\
Q(L_R^3)&=&7\,, \qquad Q(L_R^2)=6\,,\qquad Q(L_R^1)=4\,, \nonumber\\
Q(\Delta_L)&=&-7/2\,, \qquad Q(\Delta_R)=-21/2\,,\qquad Q(\phi_1)=7
\label{Qs}
\,,
\eeq
with $Q(L_i^j)=Q(L_i^{\prime j})$.
Note that the charges of $\Delta_{L,R}$ are half integers and therefore they carry an odd
parity under the residual, $Z_2$, horizontal symmetry.
All the other fields have integer charges and therefore carry an even parity.
Thus, in the limit where the discrete symmetry is exact ($\delta\to 0$) 
the Majorana mass matrices vanish. 
The discrete symmetry is assumed to be further broken by
the small parameter $\delta$.

For concreteness,  in our calculation below, 
we use the following numerical values,
\beq\label{numb}
\eps,\delta\sim 0.3\,;\ \ v_R\sim 4\TeV\ \,.
\eeq
Other combination of parameters of a similar magnitude  
yield a viable phenomenology as well.
Furthermore, for simplicity we assume
real values for the various VEVs and couplings. Consequently, we
assume no CP violation in the lepton sector.

We now arrive at a point where we can compute the
masses of the various lepton zero modes predicted by the model.

\subsection{Charged leptons}

The charged lepton mass matrix, $M^{cl}$, is read from the 
second term in the square brackets of \eq{Lcorr}. It is
given, up to order one
coefficients, by:
\beq 
M^{cl}\sim K {k_1}\pmatrix
{{\eps}^{6}&{\eps}^4& {\eps}^3\cr {\eps}^4 &{\eps}^{2} & {\eps}\cr
{\eps}^3 &{\eps} &1 } \label{Mcl} \,.
\label{MCL}
\eeq
Using the numerical value for $K {k_1}\sim 2 \GeV$, the eigenvalues
of $M^{cl}$
reproduce the required scale for the charged lepton masses (\ref{mcl}), up to
order one coefficients:
 \beq m_e\sim \eps^6
K {k_1} \sim 1~{\rm MeV}\,, \ \ m_\mu \sim \eps^2
K {k_1} \sim 150~{\rm MeV}\,,\ \ m_\tau\simeq
K {k_1}\sim 2 \GeV\,. \label{mclmodel}
\eeq 
In addition $M^{cl}$ is hierarchical and
diagonalized by, 
\beq
\theta^c_{12}\sim\eps^2\,, \qquad
\theta^c_{13}\sim\eps^3\,, \qquad  \theta^c_{23}\sim\eps
\,.
\label{clth}
\eeq

\subsection{Neutrinos}
The light neutrino
mass matrix, $M^\nu_l$, is given by:
\beq
M^\nu_l\simeq (M_\nu^D)^T (M_R^{Maj})^{-1} M_\nu^D+M_L^{Maj}
\,,
\label{Mnulight}
\eeq
with $M_\nu^D$ being the Dirac neutrino mass and $M_{R}^{Maj}$
[$M_{L}^{Maj}$]
being the Majorana mass matrix for the right [left] handed neutrinos.
The RHS of \eq{Mnulight} contains two terms. The first,
$M^\nu_{See}\equiv (M_\nu^D)^T (M_R^{Maj})^{-1} M_\nu^D$,
is related to the seesaw mechanism\cite{SeeSaw}. The second,
$M_L^{Maj}$, is induced by the VEV of $\Delta_L$.
In the following we shall calculate the elements
of each
and we shall show that the dominant contributions to $M^\nu_l$
come from $M_L^{Maj}$, while $M^\nu_{See}$ yields
small (but non-negligible) correction to $M^\nu_l$.

In our model $M_L^{Maj}$ is read from the $\Delta_L\equiv \Delta_1$ 
Yukawa interactions of \eq{Lcorr}. Since
$\Delta_{L,R}$ carry half integer charges their corresponding couplings 
are suppressed by
$\delta\times{\rm max}
\left(\eps^{|Q(\Delta_{L,R})+2Q(L_{L,R})+1/2|},
\eps^{|Q(\Delta_{L,R})+2Q(L_{L,R})-1/2|}\right)$.
Consequently, $M_L^{Maj}$ is given by,
\beq
M_L^{Maj}=v_L \delta\pmatrix{
\eps^2&1&1\cr 1&\eps & \eps^2\cr 1&\eps^2 &\eps^3 }\,,
\label{MMaj}
\eeq
where
from \eq{gammasup} we have, 
\beq
v_L\sim {k_1^2\over
v_R}\eps^{|Q(\Delta_R)-Q(\Delta_L)-2Q(\phi_1)|}={k_1^2\over
v_R}\eps^{21}
\,.
\eeq

As explained above $M^\nu_{See}$ depends on the structure of the Dirac 
neutrino mass,
$M_\nu^D$, and on the structure of the
Majorana mass matrix for the right handed neutrinos $M_R^{Maj}$.
The matrix $M_R^{Maj}$ is simply read from
$M_L^{Maj}$ with the replacement $v_L\leftrightarrow v_R$,
\beq
M_R^{Maj}=v_R \delta\pmatrix{ \eps^2&1&1\cr 1&\eps & \eps^2\cr
1&\eps^2 &\eps^3 }\,. 
\label{MRMaj}
\eeq
The matrix $M_{R,L}^{Maj}$ has an approximate $L_e-L_\mu-L_\tau$
structure~\cite{Pet82,Barb}. Thus to diagonalize $M_{R,L}^{Maj}$
requires $\theta_{12}\sim\pi/4$, $\theta_{23}\sim1$ and a very
small $\theta_{13}$ (for a recent review 
see e.g.~\cite{Rev1} and references therein).

Since the determinant of $M_{R}^{Maj}$ is given by,
\beq
{\rm Det}\left(M_{R}^{Maj}\right)\approx (v_R \delta)^3\eps\,,
\eeq 
the inverse of $M_{R}^{Maj}$  can be approximated by,
\beq
\label{MRRot} (M_{R}^{Maj})^{-1}={1\over v_R
\delta}O_{23}(\theta_{23})O_{12}(\theta_{12}) diag (1,1,1/\eps)
O_{12}^T(\theta_{12}) O_{23}^T(\theta_{23}) \,,
\eeq
where
$O_{ij}(\theta_{ij})$ is a rotation matrix on the $ij$ plane with
an angle $\theta_{ij}$ which were defined above.

The neutrino Dirac mass matrix, $M_\nu^D$, is read from the 
first term in the square brackets of \eq{Lcorr} . It is given by,
\beq
M_\nu^D\sim
K {k_1}\pmatrix
{{\eps}^{8}&{\eps}^{10}&
{\eps}^{11}\cr
{\eps}^{10}
&{\eps}^{12} &
{\eps}^{13}\cr
{\eps}^{11} &\eps^{13} &\eps^{14}
}
\label{MclD}
\,.
\eeq
We can now use eqs. (\ref{MRRot},\ref{MclD})
to calculate $M^\nu_{See}$,
\beq
M^\nu_{See}&\approx& K^{2} {\eps^{16}\over\delta}{k^2_1\over v_R}
\pmatrix
{1&{\eps}^{2}&
{\eps}^{3}\cr
{\eps}^{2}
&{\eps}^{4} &
{\eps}^{5}\cr
{\eps}^{3} &\eps^{5} &\eps^{6}
}
\pmatrix{ \eps&1&1\cr
1&1/\eps& 1/\eps\cr
1&1/\eps&1/\eps }
\pmatrix
{1&{\eps}^{2}&
{\eps}^{3}\cr
{\eps}^{2}
&{\eps}^{4} &
{\eps}^{5}\cr
{\eps}^{3} &\eps^{5} &\eps^{6}
}\nonumber\\
&\sim&\eps^{23}{k^2_1\over v_R}\pmatrix{1&\eps^2&\eps^3\cr
\eps^2&\eps^4&\eps^5\cr
\eps^3&\eps^5&\eps^6 }
\label{Msee}
\,,
\eeq
where in the second line we used the approximation,
$\eps^{6}\sim {K^2\over\delta}$.

Combining eqs. ({\ref{MMaj},\ref{Msee}) the light neutrinos mass
matrix is given by,
\beq
M^\nu_l\simeq (M_\nu^D)^T (M_R^{Maj})^{-1} M_\nu^D+M_L^{Maj}
\sim \eps^{22}{k^2\over v_R}\pmatrix{\eps&1&1\cr
1&\eps&\eps^2\cr
1&\eps^2&\eps^3}
\,,
\label{Mnulfin}
\eeq
where, as anticipated, the dominant elements come from
$M_L^{Maj}$.
Consequently, the Matrix $M^\nu_l$ has an approximate
$L_e-L_\mu-L_\tau$ structure
which therefore yields inverted hierarchical masses:
\beq
|m_1|&\simeq&|m_2|\approx\eps^{22}{k^2\over v_R} \sim0.02\eV \ ,\ \
|m_3|\approx\eps^{23}{k^2\over v_R}\sim0.007\eV\,;\nonumber \\
\Delta m^2_{\rm Sol}&\sim&2\cdot10^{-4}\eV^2 \ ,\ \ \Delta m^2_{\rm Atm}\sim 1\cdot10^{-3}\eV^2
\,,
\label{nmass}
\eeq
in agreement, up to an order one coefficients, with the recent data given in \eq{numass}.
The neutrino mixing angles read from $M^\nu_l$ and $M^{cl}$ (\ref{Mcl}) are:
\beq
\tan^2\theta_{12}&=&{\sqrt{1+\tan^2 2\theta_{12}}-1\over \sqrt{1+\tan^2
\theta_{12}}+1}
\sim {1-\eps}\sim0.7\,,\nonumber\\
\tan^2\theta_{23}&=&\Oc(1)\ ,\ \
\theta_{13}\sim\Oc(\eps^2)\sim0.1\,,
\label{mixing}
\eeq
again in agreement
with the recent neutrino data given in \eq{numix}.

\section{Testing The Model By Experiments}\label{Test}

It is very interesting to understand whether
the model has unique experimental signatures
which can be tested in near future experiments. 
Our model belongs to a class of LRS models in 5D, 
that was recently constructed in~\cite{MN,MP} and  
some of its phenomenological implications were already discussed
there. We shall briefly summarize some of the most important properties:\\
(1) Existence of Kaluza-Klein (KK) excitations of the gauge bosons;
(2) No left-right mixing in the charged sector;
(3) The lightest $W_2$ is a KK excitation while $Z_2^{(0)}$ is not;
(4) A lower bound of the order of a TeV on $L^{-1}$
and on the masses of the  $Z_2$ and $W_2$ lower KK excitations;
(5) Existence of a heavy stable lepton whose mass
is of the order of $L^{-1}$.

In addition we list below predictions which are specifically 
related to our model:
\begin{itemize}
\item[(i)] Inverted mass hierarchy for the neutrinos.
\item[(ii)] Order one mixing for $\theta_{23}$ but not
parametrically close to maximal.
\item[(iii)] Small branching ratio (BR) of lepton flavor processes such
as $\mu\to e\gamma$, $\tau\to \mu\gamma$ and
$\tau\to e\gamma$ (for details on these processes see e.g.~\cite{Rev3}
and references therein) due to the smallness of the LR mixing in the 
model~\cite{MN,MP}. The same feature is also shared by the additional 
neutral and charged scalars introduced in our model. 
This is due to their corresponding transformation
under the orbifold discrete group (\ref{transcal}) and the smallness of 
$v_L/(k_1,v_R)$~(\ref{gammasup}). 
\item[(iv)] The process $\mu^-\to e^+ e^- e^-$ 
does not require LR mixing  and therefore might be significantly
enhanced 
(similar $\tau$ decay modes have a smaller experimental
sensitivity~\cite{PDG}). 
It can be efficiently mediated by the Yukawa couplings of 
$\delta_L^{--}$ to a pair of charged lepton zero modes.
The rate is given by (see e.g. \cite{Rev3,Rev2,Moh} and references therein)
\beq\label{rate}
\Gamma(\mu\to3e)={G^2_{eff} m_\mu^5\over 192 \pi^3}\,,
\eeq
where $G_{eff}$ is proportional to the product of
($\delta_L^{--}ee$) and ($\delta_L^{--}e\mu$) Yukawa couplings. 
In our model it is given by
\beq
G_{eff}\approx{\eps^2\delta^2\over M_* R M^2_{\delta_L^{--}}}
\lsim{\eps^2\delta^2\over a^2 M^2_{\delta_L^{--}}}\sim 10^{-11}\GeV^{-2}
\label{Gef}\,.
\eeq
Comparing the above result with the 
current experimental bound~\cite{PDG},
\beq
G_{eff}\lsim 10^{-11}\GeV^{-2}\,,
\eeq
and we learn that the two are comparable. This implies
that our model will be subject to an experimental test in the very near
future.
\item[(v)] As $M^\nu_l$ (\ref{Mnulfin}) is induced by an approximate
$L_e-L_\mu-L_\tau$ symmetry,
the model predicts a rather small amplitude for neutrinoless
double decay (see e.g.~\cite{Pet1,dbd} and references therein):
\beq
|\langle m \rangle|\sim\eps^{23}{k^2\over v_R}\sim0.007\eV
\label{dbd}
\,.
\eeq
This is comparable with the lower limit for the case of inverted 
hierarchy~\cite{Pet1} and very likely, below the sensitivity of
near future experiments.
\end{itemize}

The above results are based on the assumption that the widths of the
lepton wave functions are universal.
If the Yukawa couplings between the leptons and $\varphi$ are not universal
but of order unity the model looses some of its
predictive power. 
This is mainly due to the fact the structure of the
Dirac mass matrices (\ref{MCL},\ref{MclD})
may be significantly modified due to different overlaps of the wave
functions in the extra dimension (even if the 
eigenvalues of these matrices are unchanged).
In this case the resultant mixing angles and neutrino masses
are functions of the widths and cannot be explicitly calculated. 
Note, however, that if the hierarchical structure (\ref{MCL},\ref{MclD})
is not badly broken (with the corresponding eigenvalues
remain roughly the same), 
most of the above features of the model are still valid.
This is due to the fact that the structure of the 
Majorana mass matrices (\ref{MMaj},\ref{MRMaj})  
is independent of the universality assumption and the 
neutrino masses and mixing are mainly derived from
$M_L^{Maj}$ (\ref{MMaj}).

\section{Discussion and Conclusion}\label{Conclusion}

We constructed
a left-right symmetric (LRS) model of leptons in five dimensions.
The structure of the lepton flavor parameters is explained
naturally and the gauge hierarchy problem is ameliorated.

The typical scales are: (1) The fundamental-domain size in the extra
dimension, $L\sim\TeV ^{-1}$;
(2) The LRS breaking scale, $v_R\sim 4\TeV$; 
(3) The fundamental scale, $M_*\sim 25\TeV$.

The model predicts that the light neutrino mass matrix has an
approximate $L_e-L_\mu-L_\tau$ structure
which therefore yields inverted hierarchical masses,
with $\tan^2\theta_{12}\sim 0.7$ and $\theta_{23}=\Oc(1)$
but not parametrically close to maximal.
The model predicts a relatively large branching ratio for the lepton
flavor violation process $\mu^-\to e^+ e^- e^- $, comparable with its
present experimental sensitivity.

The fact that the model copes both with the lepton
flavor puzzle and with the gauge hierarchy problem
is manifested in its rich  structure.
The model has elements related to the following mechanisms:\\
(i) The Arkani-Hamed-Schmaltz mechanism-
localization of ``left'' and ``right'' leptons on different
fixed points in the extra dimension, which then explains the smallness of
$m_\tau$ compared with the electroweak breaking scale. \\
(ii) The Froggatt-Nielsen mechanism- 
a horizontal symmetry which yields a modified see-saw mechanism and accounts
for the other flavor parameters.

The above  results are derived for the case of universal
Yukawa couplings to $\varphi$ and when $L^i_j$ and $L'^i_j$
have the same horizontal charges.
If this is not the case the model looses some of its predictive power.
It is rather clear, however, that it is possible, 
for a given set of order one Yukawa couplings, to find a charge
 assignment that
preserve the model appealing features. 
Thus, although the more generic setup of the model is less predictive,
it is still provide a natural framework to understand the origin
of the flavor parameters.

In this work we focused on the large mixing angle solution
of the solar neutrino problem as favored by the present
experimental data. In an earlier work~\cite{KP},
however, we showed that a LRSM with a similar $v_R$  
and the same horizontal symmetry, can in principle, produce a model
with the small mixing angle solution of the solar neutrino problem.

Finally we note that the addition of the quark fields
to the fermion sector of our model is in principle possible based on
the above ideas~\cite{Kap}. The large top mass, however, might require
an additional structure~\cite{GrPe}.

\acknowledgements

I thank Yossi Nir, Yael Shadmi, Oleg Khasanov and especially Yuval Grossman 
for helpful discussions and comments on the manuscript.

\newpage


\newpage
\appendix
\section{Localization of the Fermion Wave Functions}
In this part we examine the requirements for splitting 
and localizing the fermions around the different fixed
points. Some of the results derived below are known 
(see e.g. ~\cite{HS,GGH,Col}), but
for the self consistency of our work we briefly present them below. 

To localize the fermions we add a
real scalar field, $\varphi$, which transforms non-trivially under the
orbifold discrete group~\cite{HS,GGH}, 
 \begin{equation}
\varphi(x,x_5)\rightarrow \varphi(x,-x_5)=-\varphi(x,x_5)\,,
\label{phi}
\end{equation}
which for simplicity we take as a single
$Z_2$. The Lagrangian is
\begin{equation}
\cL=\bar\psi\left(i\spur\,\partial-\gamma_5\partial_5-{f\over\sqrt
M_*}\varphi\right)\psi
+{1\over2}\partial^\mu\varphi\partial_\mu\varphi
-{1\over2}\partial_5\varphi\partial_5\varphi
-\lambda\left(\varphi^2-v'^2\right)^2 \,, \label{L5}
\end{equation}
where for further simplicity we consider a single fermion model.

The shape of the fermion wave functions depends on 
the shape of $\langle\varphi\rangle$. The localization is
significant if $\int dx_5\langle\varphi(x_5)\rangle$ is large
enough. In the following we investigate what is required from the
model parameters so that efficient localization will occur.

\subsection{Properties of the profile of $\langle\varphi(x_5)\rangle$}

The boundary conditions, (\ref{phi}), require that the scalar field vanish
on the orbifold fixed points, namely, 
at $x_5=0$ and $L$. However, for $v'^2>0$,
the scalar field tends to develop a VEV.
We can make this argument more quantitative by considering 
the scalar Lagrangian, neglecting the effects of its Yukawa interactions:
\begin{equation}
\cL_\varphi={1\over2}\partial^\mu\varphi\partial_\mu\varphi
-{1\over2}\partial_5\varphi\partial_5\varphi
-\lambda\bigl(\varphi^2-v'^2\bigr)^2 \label{Lphi}\,.
\end{equation}
The minimum energy configuration
is found when the following action is minimized,
\beq S={v^2\over L^2}\int_0^1
du\left[{1\over2}\partial_5\varphi'\partial_5\varphi'
+a^2\bigl(\varphi'^2-1)^2\right]\,, \label{S} \eeq
with $u=x_5/L$,
$v=v' \sqrt{L}$, $\lambda'={\lambda M_*}$, $a^2=\lambda' vL {v
\over M_* }$ and $\varphi'={\sqrt{L}\over v}\varphi$.~\footnote{
Note that
above (\ref{scales}) we defined $a^2= vL {v \over  M_* }$. The
difference is the $\lambda'$ coefficient which is assumed to be of
order one and therefore it is omitted below.} Let us find
the condition for which $\varphi=0$ will not be a local minimum of $S$.
Expanding $\varphi$ in terms of sin functions which satisfy the
boundary condition of \eq{phi},
\beq\varphi'(u)=\sum_n \sqrt2
\varphi'_n\sin {\pi nu}\,,
\eeq 
and integrating over $u$ yields the
resultant ``mass'' matrix for the action $S$,
\beq
M^2_{nm}=\left.
{\partial^2 S\over\partial \varphi'_n \partial
\varphi'_m}\right|_{\varphi_l=0}\propto
\delta_{nm}\left({\pi^2 n^2\over2}-2a^2\right) \,. \label{Mnm} 
\eeq 
From the structure of $M_{nm}$ we learn that if
\beq a^2\simeq vL {v
\over  M_* }>{\pi^2\over4}
\label{cond} \eeq
we find negative eigenvalues 
for $M_{nm}^2$ which implies that $\varphi'=0$ is not a
local minimum of $S$~\cite{GGH}. Thus, we expect to find a lower action
$S$ for $u$ dependent $\varphi'$,
\begin{equation}
\left\langle\varphi(u)\right\rangle=h(u)\,. \label{h}
\end{equation}

An intuition on the possible profile of $h(u)$ is gained by viewing
the action $S$ as related to a one particle problem. The particle
moves in a potential $V(\varphi')=-a^2\bigl(\varphi'^2-1)^2$,
shown in figure \ref{potfig}, 
with ``initial'' and ``final''
conditions $\varphi'(u=0,1)=0$ (for more details see
e.g.~\cite{Col}). 
The solution $\varphi'(u)=0$ represents a static particle at the 
bottom of the potential.

We prove below that the ``time'' for the particle to
reach the turning point and returning is a monotonically increasing
function with the ``length'' of the corresponding ``trajectory''.
Consequently, there is only one additional solution which must correspond
to the global minimum, provided that \eq{cond} is satisfied.

Since $S$ does not contain an explicit $u$ dependence, we can
define an effective conserved ``energy'', $\tilde E$, 
\beq \tilde
E={1\over2}\partial_5\varphi'\partial_5\varphi'-a^2\bigl(\varphi'^2-1)^2
\label{Eeff} \,, \eeq where $-a^2<\tilde E<0$. 
Using this
we can estimate the maximal value of $\varphi'$ (the value at the
``turning point'') and study its dependence on the time period,
$\tilde T$:
\beq {\tilde T\over 4}
=1/2&=&{1\over\sqrt2 a}\int_0^{\varphi'(1/2)}{d\varphi'\over  \sqrt
{\tilde E/a^2-(\varphi'^2-1)^2}}
\nonumber \\
&=&{1\over\sqrt2 a}\int_0^{\varphi'(1/2)}{d\varphi'\over 
\sqrt{[\varphi'^2-\varphi'(1/2)^2][\varphi'^2-\bar\varphi'(1/2)^2]}}\equiv
I \label{turn} \,, \eeq with $\varphi'(1/2)^2=1-\sqrt{-\tilde
E/a^2}$, is the physical turning point and
$\bar\varphi'(1/2)^2=1+\sqrt{-\tilde E/a^2}=2-\varphi'(1/2)^2$ is
an unphysical one.

Using the substitution $X=\varphi'/\varphi'(1/2)$, $I$ takes the
form of a complete eliptic integral (see e.g. \cite{RH}),
\beq
I={1\over \sqrt{2a^2 \varphi'(1/2)^2 }}\int_0^1 {dX\over
\sqrt{[X^2-1][X^2-1/d^2]}}={F(\pi/2,d)\over \sqrt{2a^2
\bar\varphi'(1/2)^2 }} \label{I}\,,
\eeq with
$d^2=\varphi'(1/2)^2/\bar\varphi'(1/2)^2$ and $F(\pi/2,d)$ is a
complete elliptic integral of the first kind.

The function $F$ can be expressed as a series in $d^2$, (provided
that $0<d<1$ as in our case) \cite{RH}:
\beq F(\pi/2,d)\sim
{\pi\over2}\left\{1+\left({1!!\over 1!\cdot2^1}\right)^2
d^2+\dots+ \left[{(2n-1)!!\over n! 2^n }d^{2n}\right]^2\right\}
\label{F} \,. \eeq
As a consistency check for our results we note
that for a very small $\varphi'$ the potential is quadratic in
$\varphi'$, $V(\varphi')\sim 2a^2\varphi'^2$. This implies that
the time period should be given by \beq \tilde T=\pi/a
\,.\label{check}\eeq Plugging the first term in the expansion of
$F(\pi/2,d)$ into the expression for $\tilde T$ in \eq{turn}
indeed reproduces the result of \eq{check}.

We are now at a point were we can show that the ``time'' for the
particle to reach the turning point and returning is a monotonically
increasing function with the ``length'' of the corresponding
``trajectory'' or ``energy''. Since all the coefficients of the
series in \eq{F} are positive and $d$ is monotonic with $\tilde E$
(in the physical range) it is clear that  $F(\pi/2,d)$ is a
monotonic increasing function with   $\tilde E$. The time period
is given by $I\propto F(\pi/2,d)/\bar\varphi'(1/2)$. Since
$\bar\varphi'(1/2)$ decreases with $\tilde E$, $I$ or the time
period is indeed increasing with the energy which guarantees that
there is only one additional minimum of $S$.

For the above potential, $V(\varphi)$,  one cannot compute
analytically the profile of  $\langle\varphi(x_5)\rangle$. However
to our consideration we only need to have information on its
asymptotical behavior.
This was analyzed in~\cite{GGH} in which it was shown that for large
$L$ the profile is approximated by, in our scaling: 
\beq h(u)\approx \tanh \left(\sqrt2au\right)\tanh
\left[\sqrt2a(1-u)\right]\label{kink}\,,\eeq
where the exact $u$ dependence of (\ref{kink}) is found by taking the large
$a$ limit of \eq{Eeff}.
With that profile we expect a significant
reduction in the overlap between the wave functions localized at
the different fixed points.

We demonstrate that this case corresponds to the
large $a$ case using our semi-analytic derivation presented above. 
In addition we gain physical understanding  for the resultant shape of $h(u)$.
Consider the result of eqs. (\ref{turn},\ref{I}) when $d\sim1$ or
$\varphi'(1/2)^2\sim 1-\eps^2$. In that case $F(\pi/2,d)$ can be
approximated by the following expression~\cite{RH}: 
\beq
F(\pi/2,d)\sim \ln4-\ln\sqrt2\eps \label{F2} \,. 
\eeq 
Substituting
this into \eq{turn} yields the following equation for
$\varphi'(1/2)\,$: 
\beq a/\sqrt2=\ln4-\ln\sqrt2\eps \label{turn2} 
\,,
\eeq
which implies 
\beq \varphi'(1/2)^2=1-8e^{-a\sqrt2} \label{turn22}
\,.
\eeq 
From (\ref{turn22}) we learn that indeed this limit is good
for a relatively large $a$. This means that at $\varphi'\sim0$ the
kinetic energy is very large and there is a very rapid motion
there. Furthermore, since near the turning point the potential is
almost flat the net force on the particle is very small. This
implies that the particle spends most of its time near the turning
point moving with a nearly constant velocity. Thus, as promised,
$h(u)$ can be well approximated by the two kink profile given in
\eq{kink}. We also learn from (\ref{turn22}) that this limit is
good for rather modest values of $a$. For example $a\geq4$ yields
$\varphi'(1/2)^2> 0.97$.

\subsection{The shape of the fermion wave functions}
We analyzed above the profile of 
$\langle \varphi(x_5)\rangle$, relevant for our considerations. 
In this part we ask what are the conditions for
the fermions to be localized around the different fixed points with
small overlap of the corresponding wave functions.
Using the derivation of \cite{HS,GGH}
we find that the shape of a zero mode fermions in the fifth
dimension is given by:
\beq
\xi_{0+}(x_5)=K_{\xi_{0+}} e^{-s(x_5)}
\label{xipl}
\eeq
with $K_{\xi_{0+}}$ being a normalization constant and
\beq
s(x_5)=f{\sqrt L v \over \sqrt M_*}\int_0^{x_5\over L}du\, h(u)=
f a \int_0^{x_5\over L}du\, h(u)
\label{s}\,,
\eeq
where, as shown in \cite{GGH}, if $f$ is positive [negative] the
corresponding fermion will be localized around $x_5=0\,[L]$.
As explained above the fact that $\varphi$ is odd under the LRS
discrete group (\ref{GLR}) guarantees that the 
fields $L_{1},L'_{1}$ [$L_{2},L'_{2}$] 
have positive [negative] Yukawa coupling $f$.

In order to estimate the overlap between the wave functions we
approximate $h(u)$ by a step function~\cite{Kap},
$h(u)\sim 1$, in the required range. Consequently, the wave functions 
of two fermions with opposite Yukawas 
are roughly given by
\beq
\psi_{1,2}\approx K_{1,2}e^{\pm |f_{1,2}|a u}
\,.
\eeq 
Thus, the overlap between the wave functions is
given, up to an order one coefficients, by~\cite{Kap}:
\beq K\equiv \int_0^1 du\psi_1\psi_2
\sim
e^{-Ca}
\label{K}\,,
\eeq
where $C$ is an order one coefficient, that for simplicity, we assume to
be one.
In the range $a=5..9$ we get that the overlap between the 
wave functions is indeed small, as required by our model, 
of the order of $m_\tau/k_1$.

\putFig{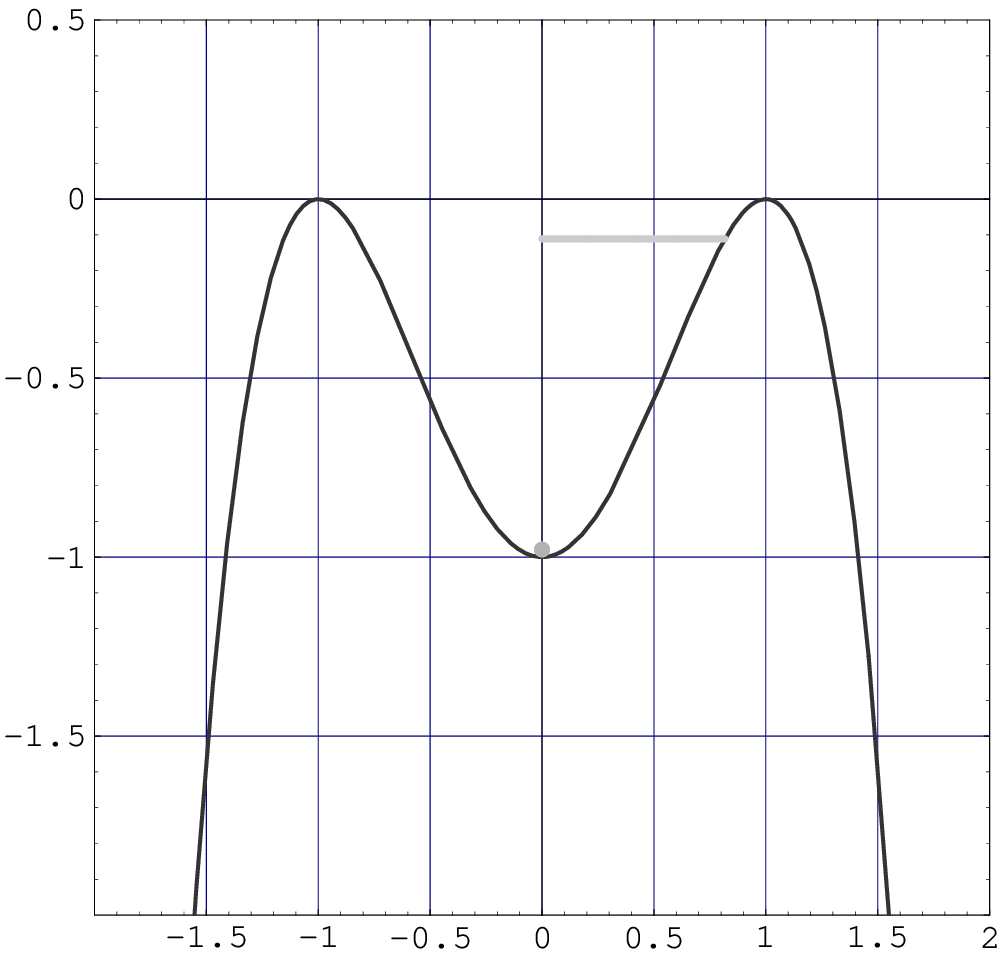}{The black line corresponds to the effective
potential which determines $h(x_5)$. The horizontal
gray line represents a typical ``energy''  of
$h(x_5)$, the ``trajectory'' which minimizes the effective action
$S$ of \eq{S}. The gray point, at $\left[\varphi',{V(\varphi')\over
a^2}\right]=[0,-1]$, corresponds to the ``static'' solution.}
{\widthAR}{\heightAR}{\spaAR} {\varphi'}{{V(\varphi')\over a^2}}


\end{document}